\documentclass[apj]{emulateapj}

\def\ltsim{ \,{}^<_\sim\, }
\def\gtsim{ \,{}^>_\sim\, }
\shorttitle{Radio study of the poor cluster AWM4}
\shortauthors{S.~Giacintucci et al.}
\begin{document}

\title{A GMRT multifrequency radio study of the isothermal core of the poor galaxy cluster AWM\,4}

\author{Simona Giacintucci\altaffilmark{1,2}, 
Jan M. Vrtilek\altaffilmark{2},
Matteo Murgia\altaffilmark{1,3},
Somak Raychaudhury\altaffilmark{2,4},
Ewan J. O'Sullivan\altaffilmark{2},
Tiziana Venturi\altaffilmark{1},
Laurence P. David\altaffilmark{2},
Pasquale Mazzotta\altaffilmark{2,5},
Tracy E. Clarke\altaffilmark{6,7},
Ramana M. Athreya\altaffilmark{8}
}
\altaffiltext{1}{INAF -- Istituto di Radioastronomia, 
via Gobetti 101, I-40129, Bologna, Italy; sgiaci$_{-}$s@ira.inaf.it}
\altaffiltext{2}{Harvard-Smithsonian Center for Astrophysics, 
60 Garden Street, MS-67, Cambridge, MA 02138, USA}
\altaffiltext{3}{INAF -- Osservatorio Astronomico di Cagliari, Loc.
Poggio dei Pini, Strada 54, I--09012 Capoterra, Italy}
\altaffiltext{4}{School of Physics \& Astronomy, University of
Birmingham, Birmingham B15 2TT, United Kingdom}
\altaffiltext{5}{Dipartimento di Fisica, Universit\'a di Roma Tor Vergata,
via della Ricerca Scientifica 1, I--00133, Roma, Italy}
\altaffiltext{6}{Naval Research Laboratory, Washington, DC. 20375, USA}
\altaffiltext{7}{Interferometrics, Inc., 13454 Sunrise Valley Dr. No. 240, Herndon, VA 20171, USA}
\altaffiltext{8}{Tata Institute of Fundamental Research, National Centre for
Radio Astrophysics, Ganeshkhind, Pune, 411 007, India}

\begin{abstract}
We present a detailed radio morphological study and spectral analysis
of the wide--angle--tail radio source 4C\,+24.36 associated with the
dominant galaxy in the relaxed galaxy cluster AWM\,4. Our study is
based on new high sensitivity GMRT observations at 235 MHz, 327 MHz 
and 610 MHz, and on literature and archival data at other 
frequencies. We find that the source major axis is likely oriented 
at a small angle with respect to the plane of the sky. The 
wide--angle--tail morphology can be reasonably explained by adopting 
a simple hydrodynamical model in which both ram pressure (driven by 
the motion of the host galaxy) and buoyancy forces contribute 
to bend the radio structure. The spectral index progressively steepens 
along the source major axis from $\alpha \sim$0.3 in the region close to 
the radio nucleus to beyond 1.5 in the lobes. The results of the analysis 
of the spectral index image allow us to derive an estimate of the radiative 
age of the source of $\sim$ 160 Myr. The cluster X--ray emitting gas has 
a relaxed morphology and short cooling time, but 
its temperature profile is isothermal out to at least 160~kpc from the 
centre. Therefore we seek evidence of energy ejection from the central 
AGN to prevent catastrophic cooling. We find that the energy injected by 
4C\,+24.36 in the form of synchrotron luminosity during its lifetime is 
far less than the energy required to maintain the high gas temperature in 
the core. We also find that it is not possible for the central source to 
eject the requisite energy in the intracluster gas in terms of the enthalpy 
of buoyant bubbles of relativistic fluid, without creating discernible large 
cavities in the existing X--ray {\it XMM--Newton} observations. 

\end{abstract}

\keywords{galaxies: clusters: general --- galaxies: clusters: individual (AWM\,4) 
--- intergalactic medium --- radio continuum: galaxies --- X--rays: galaxies: clusters}

\section{Introduction}\label{sec:intro}

In the cores of relaxed clusters and groups of galaxies the cooling 
time for the hot X--ray emitting intracluster medium (ICM) can be 
substantially less than the Hubble time (e.g., Edge et al. 1992; 
Sanderson et al. 2006). In the absence of any substantial sources of 
heating, the gas in these regions will cool and recombine, constituting 
a cooling flow (Fabian \& Nulsen 1977; Cowie \& Binney 1977). However, 
high resolution X--ray imaging and spectroscopy, using the {\it Chandra} 
and {\it XMM--Newton} observatories, reveal unusually low quantities of 
cooling gas compared to the expectations from the standard cooling flow 
model (e.g., Fabian 1994), showing that, even in the cores, the gas temperature 
does not fall below $\sim$1--2 keV (e.g., Peterson et al. 2003; Kaastra et 
al. 2004). The fate of the cooling gas, and the possible involvement of 
additional heating ({\it feedback}) to compensate for the observed radiative 
losses and regulate the cooling by (re--)heating the gas, remains the 
subject of an extensive debate.

One of the most promising sources of feedback is the active galactic 
nucleus (AGN) harbored in the central dominant galaxy in these systems 
(e.g., Tabor \& Binney 1993; Nusser et al. 2006; Nulsen et al. 2006; 
see also the recent review by McNamara \& Nulsen 2007, and references 
therein). Indeed, radio and X--ray images provide direct evidence of 
the widespread existence of AGN--driven phenomena in an increasing number 
of cool core clusters and groups. Evidence of strong interactions between 
the radio and thermal plasma can be found in the detection of X--ray 
disturbances in the hot gas, such as cavities, edges and filaments, 
which appear correlated with the structure of the central radio galaxy 
(see for instance the review by Blanton 2004). However, the mechanism of 
the energy transfer from the radio source to the surrounding thermal gas 
(e.g., shocks, gravity waves, turbulence) is still unclear, as is the 
mechanism of the distribution of energy throughout the cooling region 
(e.g., McNamara \& Nulsen 2007). 
\\
\\
The combination of high quality X--ray imaging and high sensitivity 
radio observations is a powerful tool to investigate the ICM/AGN 
connection in clusters and groups. In particular, the examination 
of radio images at multiple frequencies (especially at frequencies 
$\le$1~GHz) is important to study the cycle of radio activity and 
elucidate the timescales and physical mechanisms of the energy 
injection. With this aim in mind we recently have embarked upon a 
low radio--frequency survey of a sample of groups of galaxies with 
the Giant Metrewave Radio Telescope (GMRT). High quality X--ray 
images from {\it Chandra} and/or {\it XMM--Newton} observations show 
that most of these systems have a structure in the X--ray suggestive 
of strong interaction between the central radio source and the intragroup 
gas (Giacintucci et al. in preparation). The survey aims to investigate 
the radio source properties over a broad frequency range, examine 
the effects of the AGN at various phases of its activity, and study 
the geometry, timescales and physical mechanisms of the energy injection.

In this paper we present new GMRT radio images, obtained at 235 MHz, 
327 MHz and 610 MHz of the poor cluster of galaxies AWM\,4, which belongs 
to the sample selected for the survey. The cluster hosts the radio source 
4C\,+24.36, which is associated with the central giant elliptical galaxy 
NGC\,6051. The general properties of the system are summarised in 
Tab.~1, where we provide the following information: J2000 coordinates, 
redshift, absolute R magnitude of NGC\,6051; 1.4 GHz flux density of 
4C\,+24.36 from the NRAO VLA Sky Survey (NVSS; Condon et al. 1998), 
and the corresponding radio power calculated at the redshift of the optical galaxy\footnote{Throughout the paper we assume H$_0$ = 70 km sec$^{-1}$ Mpc$^{-1}$, 
$\Omega_m$ = 0.3, and $\Omega_{\Lambda}$ = 0.7.}; we also list the cluster 
optical velocity dispersion from Koranyi \& Geller (2002), and 
emission--weighted gas temperature of the ICM from O'Sullivan et al. 
(2005; hereinafter OS05). The linear scale, at the redshift of NGC\,6051 
given by the cosmology adopted in this paper, is 0.624 kpc per arcsec.

The 1.4 GHz radio power of 4C\,+24.36 and absolute magnitude of the host 
galaxy (Tab.~1) place the source on the lower end of the FRI--FRII transition 
region (e.g., Owen \& Ledlow 1994). The source is extended at the 
5$^{\prime \prime}$ resolution and sensitivity (0.15 mJy b$^{-1}$) of the 
1.4 GHz image from the FIRST survey (Becker et al. 1995), and shows curved 
jets traceable up to $\sim$50 kpc from the nucleus along the overall direction 
of the galaxy minor axis.

%
%

\begin{table}[h!]
\label{tab:prop}
\caption[]{General properties of the AWM\,4 system.}
\begin{center}
\begin{tabular}{llc}
\hline\noalign{\smallskip}
NGC\,6051 & & \\
& RA$_{J2000}$ (h,m,s) & 16 04 56.8 \\
& DEC$_{J2000}$ ($^{\circ}$, $^{\prime}$, $^{\prime \prime}$) & +23 55 56 \\
& z & 0.0312 \\
& M$_{\rm R}$ & $-$23.01 \\
&& \\
4C\,+24.36 & & \\
& S$_{\rm 1.4\,GHz, \, NVSS}$ (mJy) & 608 \\
& logP$_{\rm 1.4\,GHz}$ (W Hz$^{-1}$) & 24.15 \\
&& \\
AWM\,4 && \\
& $\sigma_{\rm v}$ (km s$^{-1}$) & 400 \tablenotemark{(a)}\\
& kT (keV) & 2.5 \tablenotemark{(b)}\\
&& \\
& linear scale & 0.624 kpc$ / ^{\prime \prime}$ \\
\noalign{\smallskip}
\hline
\end{tabular}
\end{center}
\tablenotetext{(a)}{Koranyi \& Geller (2002);} 
\tablenotetext{(b)}{O'Sullivan et al. (2005).}
\end{table}


The outline of the paper is as follows: in Sec.~\ref{sec:obs} 
we describe the GMRT radio observations and data reduction; 
the radio images of 4C\,+24.36 are presented in Sec.~\ref{sec:images}; 
in Sec.~\ref{sec:spectra} we study the integrated radio spectrum and 
spectral index image of the source, and derive its physical parameters; 
the results of the radio analysis are discussed in Sec.~\ref{sec:disc}; 
in Sec.~\ref{sec:xray} we discuss the X--ray properties of the 
cluster environment in the context of its radio properties. 
The summary and conclusions are given in Sec.~\ref{sec:summary}.

\section{GMRT observations and data reduction}
\label{sec:obs}


\begin{table*}[h!]
\caption[]{Details of the GMRT observations of 4C\,+24.36}
\begin{center}
\begin{tabular}{ccccccc}
\hline\noalign{\smallskip}
$\nu$ &  $\Delta \nu$ &  Date  & Observation  & HPBW, PA  &   rms      \\ 
 (MHz)  &   (MHz)         &          &  time (min)      &
	    ($^{\prime \prime} \times^{\prime \prime}$, $^{\circ}$) & (mJy b$^{-1}$) \\
\noalign{\smallskip}
\hline\noalign{\smallskip}
235   & 16 & 2006 Jul 08 & 120 & 12.7$\times$10.4, 75  & 0.80 \\
327   & 32 & 2006 Jun 16 & 100 & 9.0$\times$7.8, 53 & 0.40 \\
610   & 32 & 2006 Aug 24 & 160 & 5.0$\times$4.0, 43 & 0.05 \\
\hline{\smallskip}
\end{tabular}
\end{center}
\label{tab:obs}
\end{table*}


\begin{figure*}[h!]
\centering
\plotone{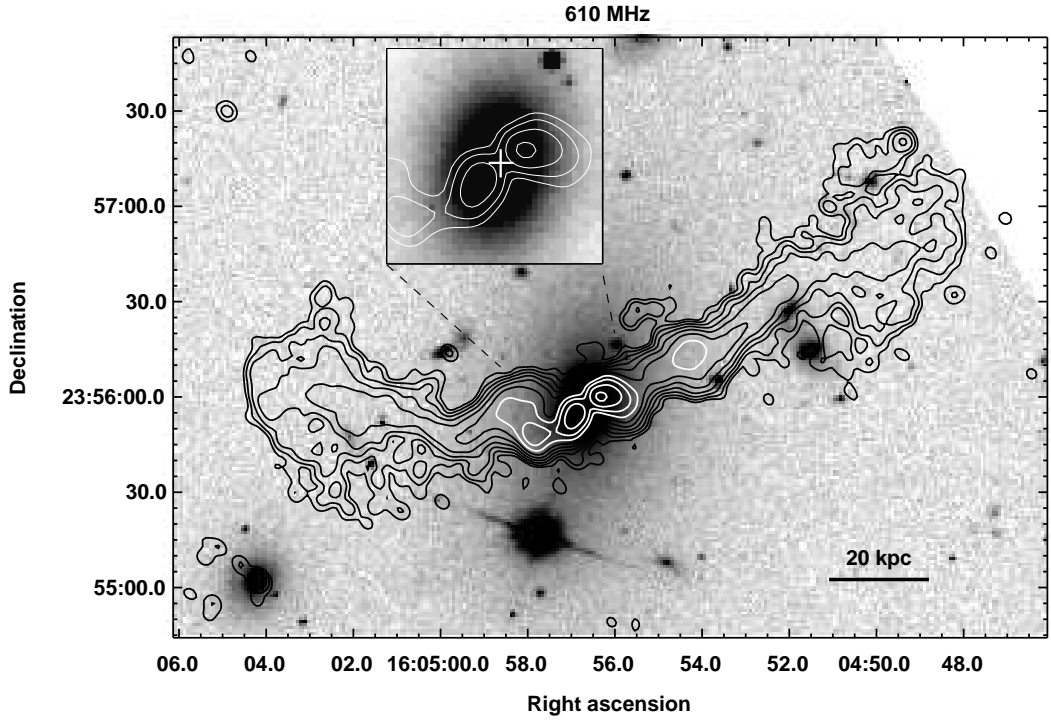}
\caption{GMRT 610 MHz radio contours of 4C\,+24.36, 
overlaid on the red optical image from the Sloan Digital 
Sky Survey. The 1$\sigma$ level in the radio image is 
50 $\mu$Jy b$^{-1}$. The contour levels are 0.2, 0.4, 
0.8, 1.6, 3.2, 6.4 (black), 12.8, 25.6, 51.2 and 102.4 
(white) mJy b$^{-1}$. The contour peak flux is 121.7 mJy 
b$^{-1}$. The restoring beam is $5.0^{\prime \prime} 
\times 4.0^{\prime \prime}$, p.a. 43$^{\circ}$. The white
cross in the insert shows the position of the radio core (see
Fig.~3).}
\label{fig:hr_image610}
\end{figure*}



\begin{figure*}[h!]
\centering
\plottwo{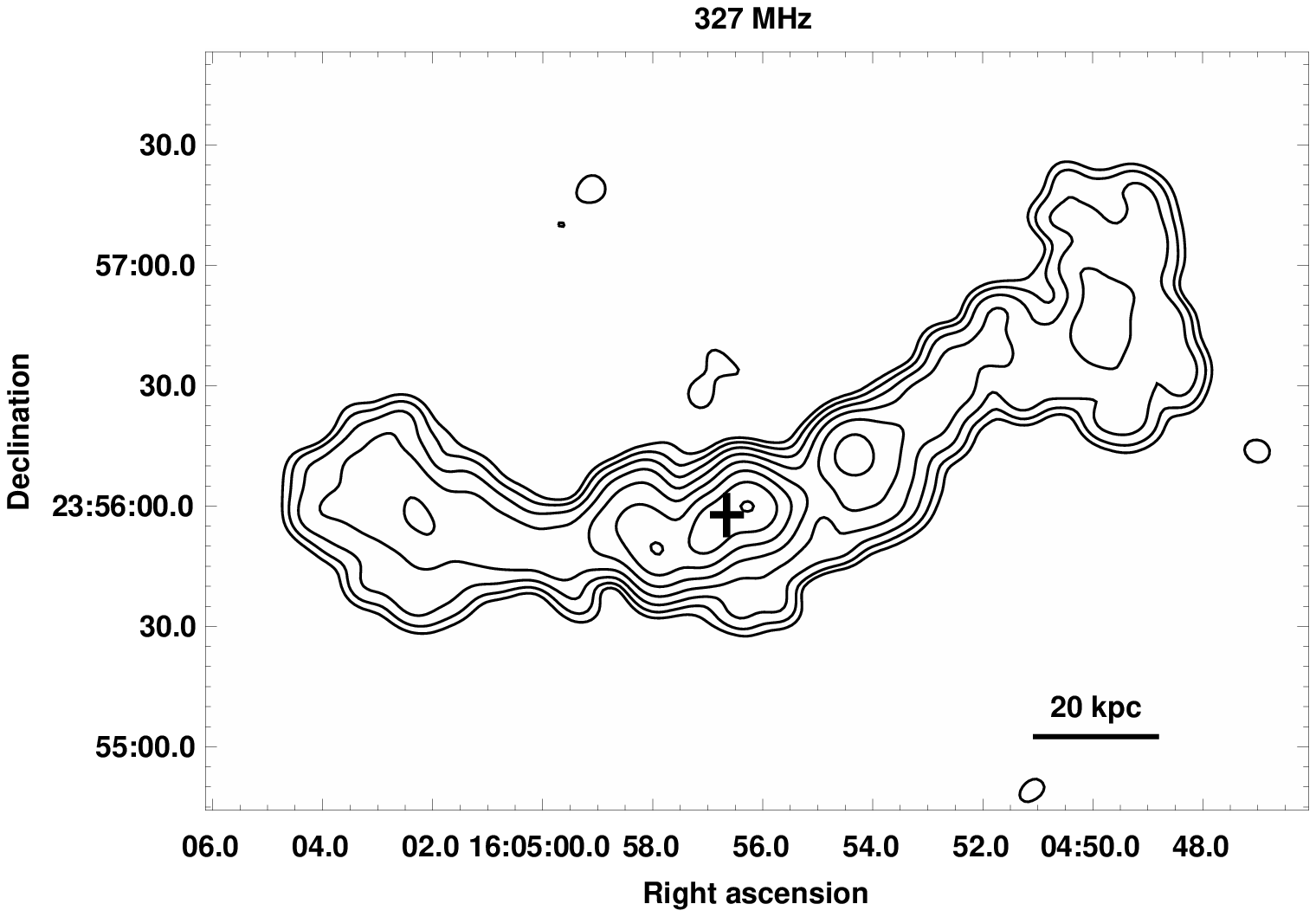}{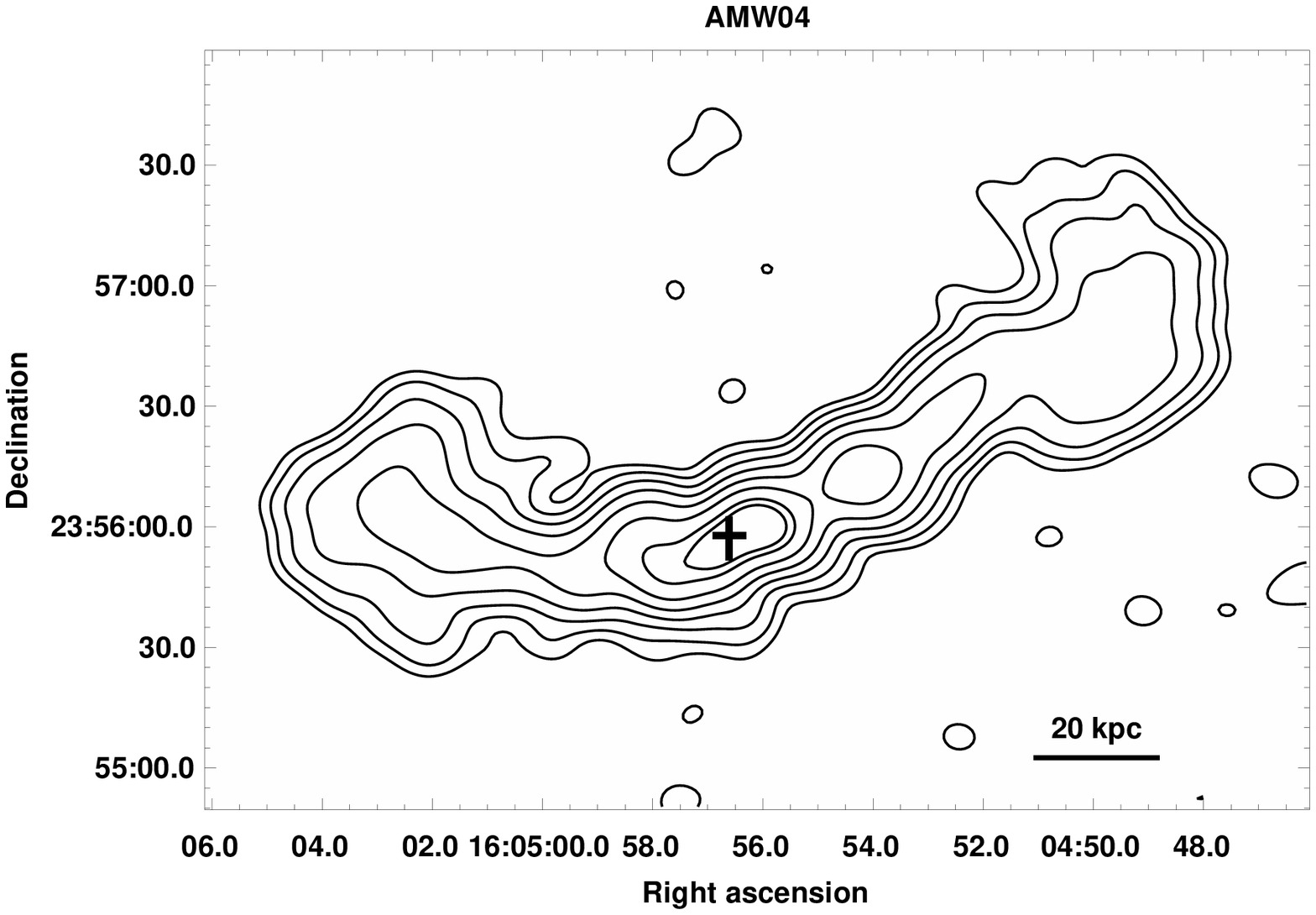}
\caption{GMRT radio contours at 327 MHz ({\it left panel}) and 
235 MHz ({\it right panel}) of 4C\,+24.36. The 1$\sigma$ level
in the image is 0.4 mJy b$^{-1}$ at 327 MHz and 0.8 mJy b$^{-1}$
at 235 MHz. Contour levels start from $\pm$3$\sigma$, and scale 
by a factor 2. The contour peak flux is 199.8 mJy b$^{-1}$ at 
327 MHz and 427.8 mJy b$^{-1}$ at 235 MHz. The restoring beam 
is $9.0^{\prime \prime} \times 7.8^{\prime \prime}$, p.a. 
53$^{\circ}$ in the left panel and $12.7^{\prime \prime} 
\times 10.4^{\prime \prime}$, p.a. 75$^{\circ}$ in the right 
panel. The black cross in both images indicates the position
of the radio core (see Fig.~3).} 
\label{fig:hr_images}
\end{figure*}


We observed the radio source 4C\,+24.36 using the GMRT at the 
frequencies of 235 MHz, 327 MHz and 610 MHz. We summarise the 
details of the observations in Tab.~\ref{tab:obs}, where we provide 
the frequency and total bandwidth, observation date, total time 
on source, half power beamwidth (HPBW) of the full array, and rms level 
(1$\sigma$) in the full resolution image. The observations were performed 
using both the upper and lower side band (USB and LSB) at each frequency, 
for a total observing band of 32 MHz at 327 and 610 MHz, and 16 MHz at 235 
MHz (Tab.~2). The data were collected in spectral--line mode with 128 
channels/band at 327 and 610 MHz, and 64 channels/band at 235 MHz 
(the spectral resolution is 125 kHz/channel). At all frequencies, the USB 
and LSB datasets were calibrated and reduced individually using the NRAO 
Astronomical Image Processing System (AIPS) package. The 
flux density scale was set using the sources 3C\,286 and 3C\,48 
as amplitude calibrators and the Baars et al. (1977) coefficients.
Careful editing was necessary to identify and remove the data affected 
by radio frequency interference at 235 and 327 MHz.
\\
In order to find a compromise between the size of the dataset and 
the need to minimize bandwidth smearing effects within the primary 
beam, the central channels in each individual dataset were averaged 
to 6 channels of $\sim$1 MHz each at 235 MHz, and $\sim$2 MHz each at 
327 and 610 MHz after bandpass calibration. At each step of the data 
reduction we implemented the standard wide--field imaging technique. 
After a number of phase self--calibration cycles, the final USB and 
LSB datasets were further averaged from 6 channels to 1 single channel 
and then combined together to produce the final images of the source. 
All images were corrected for the primary beam attenuation.
The residual amplitude errors are of the order of 
$\ltsim$ 5 \%.

\section{The radio images}\label{sec:images}

The GMRT 610 MHz image of 4C\,+24.36 is presented in 
Fig.~\ref{fig:hr_image610}, superposed on the red optical image from 
the Sloan Digital Sky Survey. The images at 327 and 235 MHz are 
shown in the left and right panels of Fig.~\ref{fig:hr_images}, 
respectively. The radio galaxy exhibits a very similar morphology 
at all frequencies and resolution, with twin continuous jets and 
extended radio lobes slightly bent toward the North. It can be classified 
as a wide--angle--tail (WAT) source, although the angle between 
the tails is rather large, i.e. $\sim 140^{\circ}$. For comparison, the 
angle between the tails of the radio galaxy 3C\,465, which is considered 
the prototype of WAT sources, is $\sim 90^{\circ}$ (e.g., Eilek et al. 1984). 
Its total angular extent is $\sim$250$^{\prime \prime}$ along 
the South--East/North--West axis in all the images shown in 
Figs.~\ref{fig:hr_image610} and \ref{fig:hr_images}. The corresponding 
largest linear size is $\sim$ 160 kpc.


\begin{table*}[t]
\caption[]{Radio properties of 4C\,+24.36}\label{tab:table3}
\begin{center}
\begin{tabular}{lccccc}
\noalign{\smallskip}
\hline\noalign{\smallskip}
& S$_{\rm 235\,MHz}$  & S$_{\rm 327\,MHz}$ &S$_{\rm 610\,MHz}$ & $\alpha_{\rm 235\,MHz}^{\rm 610\,GHz}$ & LS \\ 
\noalign{\smallskip}
& (Jy) & (Jy)  &  (Jy) & ($\pm 0.03$) &  (kpc $\times$ kpc) \\
\hline\noalign{\bigskip}			      
Total source & 2.75$\pm$0.14 & 2.20$\pm$0.11  & 1.45$\pm$0.07  & 0.67 & 160 $\times$ 40  \\
&&&&& \\
Jet West &  1.09$\pm$0.05 & 0.92$\pm$0.05 & 0.62$\pm$0.03 & 0.59  & 45$\times$15 \\
Jet East &  1.01$\pm$0.05 & 0.82$\pm$0.04 & 0.60$\pm$0.03 & 0.55  & 35$\times$15 \\
&&&&& \\
Lobe West & 0.32$\pm$0.02 & 0.23$\pm$0.01 & 0.12$\pm$0.01 & 1.03  & 40$\times$36 \\
Lobe East & 0.33$\pm$0.02 & 0.24$\pm$0.01 & 0.11$\pm$0.01 & 1.15  & 40$\times$36 \\    
\noalign{\smallskip}
\hline\noalign{\smallskip}
\end{tabular}
\end{center}
\label{tab:source}
\end{table*}



\begin{figure*}
\centering
\plotone{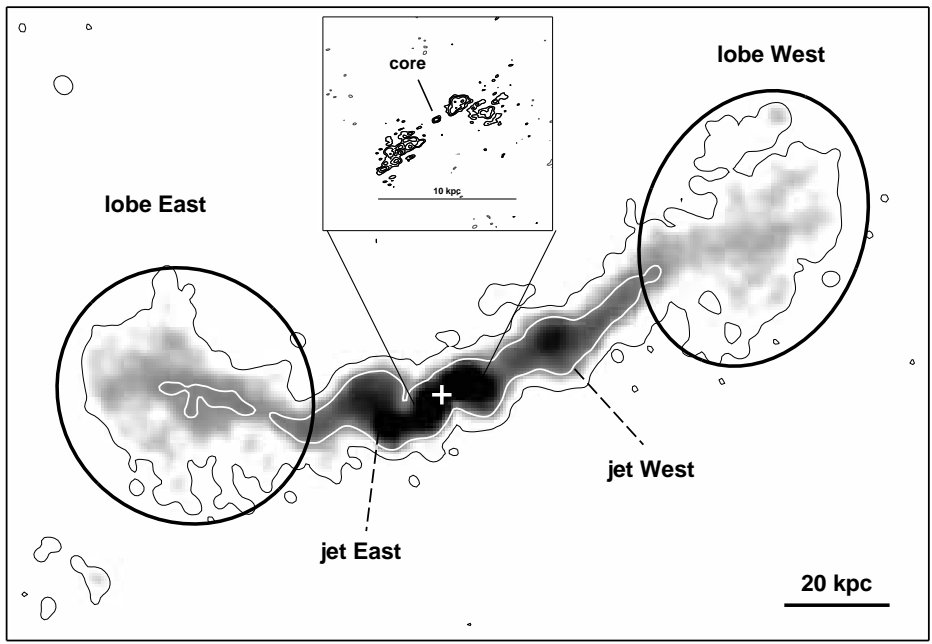}
\caption{Radio components of 4C\,+24.36. The grey scale image 
is the 610 MHz image (same as contour image in 
Fig.~\ref{fig:hr_image610}). 
The black contour level is 0.2 mJy b$^{-1}$, the white 
contour is 0.4 mJy b$^{-1}$. The white cross indicates 
the position of the radio core, as determined from  
the 0.7$^{\prime \prime} \times 0.4^{\prime \prime}$
image at 4.9 GHz (from VLA archival observations), 
shown in the insert (contours are spaced by a factor 2 starting
from $\pm$0.15 mJy b$^{-1}$). The white cross in the insert 
indicates the position of the radio core (see Fig.~3).}
\label{fig:cartoon}
\end{figure*}


Figure \ref{fig:cartoon} sketches the most prominent components of the 
source, i.e. the two bright well--collimated jets (labelled as jet West 
and jet East), extending in opposite directions with respect to the position 
of the radio core (white cross), and two roughly round and very extended 
radio lobes (lobe West and lobe East). Since the core of the radio source is 
undetected in our GMRT images (Figs.~\ref{fig:hr_image610} and 
\ref{fig:hr_images}), its position has been determined using a 4.9 GHz image 
at the resolution of 0.7$^{\prime \prime} \times 0.4^{\prime \prime}$, 
obtained from a new reduction of archival Very Large Array (VLA) data 
(Obs. Id. AK360). The 4.9 GHz image is shown as insert in Fig.~\ref{fig:cartoon}.

The most relevant properties of the source and its components, derived 
from the observations presented here, are summarised in Table \ref{tab:source}, 
where we provide the flux density at 235, 327 and 610 MHz, the spectral index 
in the 235--610 MHz frequency range (defined according to S$\propto 
\nu^{-\alpha}$), and the linear size (LS). It is clear from 
Figs.~\ref{fig:hr_images} and \ref{fig:cartoon} and Tab.~\ref{tab:source}, 
that both the jets and both the lobes are rather symmetric in total 
flux density, spectral index and extent. In particular, the western jet has 
a flux density which is only $\sim 3-10$ \% higher (depending on the frequency) 
than the eastern jet, while the lobe West is $\sim$3--8 \% fainter than the 
other. The high symmetry observed in 4C\,+24.36 suggests that the radio source 
major axis lies close to the plane of the sky.

\subsection{The radio jets}\label{sec:jets}

The 610 MHz image of 4C\,+24.36 (Figs.~\ref{fig:hr_image610} 
and \ref{fig:cartoon}) shows that the radio jets are 
characterized by a strikingly complex morphology,
with a number of wiggles, almost symmetrically placed on 
either sides within a projected distance of $\sim$ 25 kpc 
from the core.

\begin{figure*}
\centering
\plotone{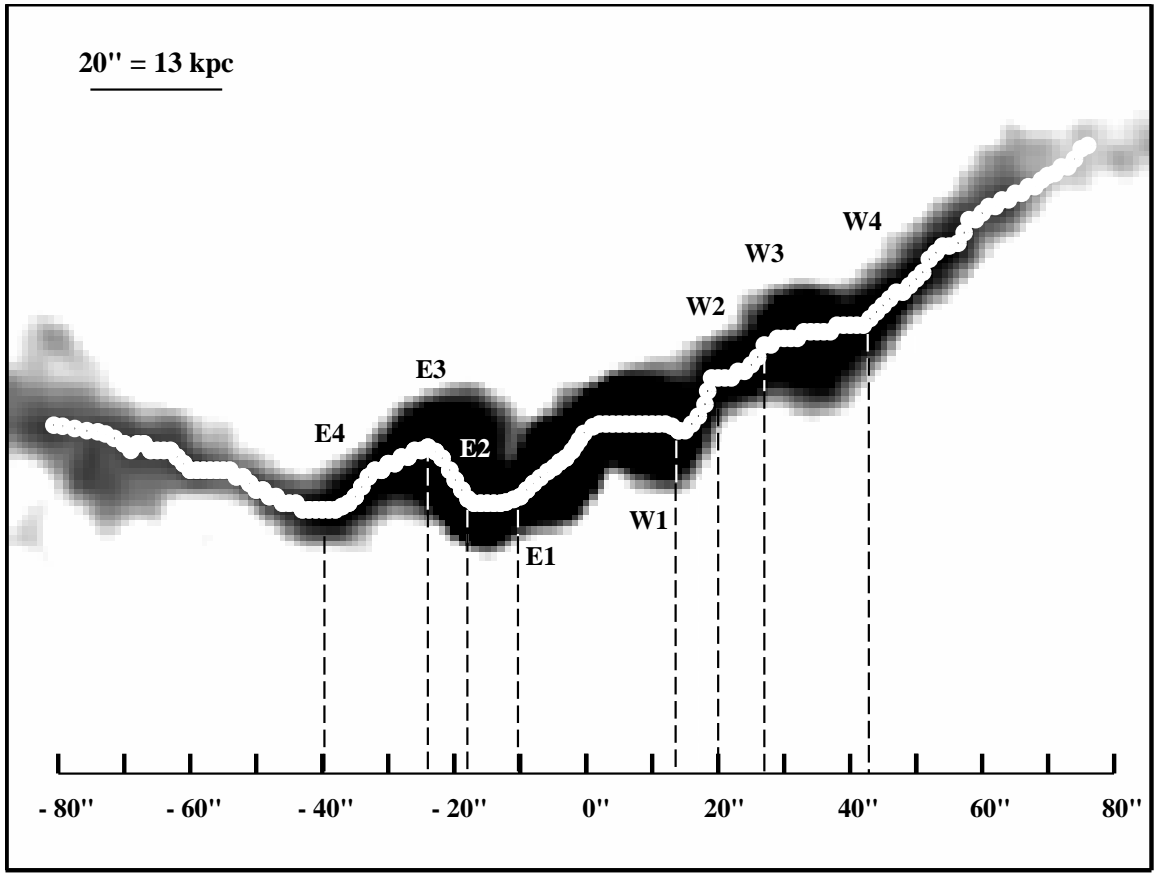}
\caption{Projected distribution of the radio surface 
brightness peaks at 610 MHz along the jets of 4C\,+24.36. 
The position of the jet bends are labelled from W1 to W4 
for the jet West, and from E1 to E4 for the jet East. 
0$^{\prime \prime}$ corresponds to the position of the radio 
core.}
\label{fig:profile}
\end{figure*}

In order to analyze the jet behaviour, we constructed 
a grid of 1$^{\prime \prime}$ wide strips, parallel 
to each other and transverse to the source major axis.
The strips were set in order to cover the whole jet 
region on the 610 MHz image (Fig.~\ref{fig:hr_image610}). 
We determined the position of the surface brightness 
peak in each stripe, and used it to reconstruct the jet 
{\it trajectory} (projected on the plane of the sky) 
given in Fig.~\ref{fig:profile}. The figure shows that 
both jets initially emerge from the core region, bent 
with a relative angle of $\sim$130$^{\circ}$. Going outward 
they undergo four prominent sharp bends (labelled W1 to W4 for 
the western jet, and E1 to E4 for the eastern jet), which 
occur at similar projected distances from the galaxy nucleus. 
In particular, the first two bends E1 and W1 are observed 
at $\sim$7 kpc from the core, W2 and E2 occur at a distance 
of $\sim$ 13 kpc, while W3 and E3 take place at $\sim$ 16 kpc.
After the last change of direction in W4 and E4 (at $\sim$ 
25 kpc from the center), the jets travel roughly straight 
for a while, and finally flare into the diffuse lobes.


\subsection{The radio lobes}\label{sec:lobes}

Both radio lobes of 4C\,+24.36 show a regular and 
roughly round morphology, and lack the presence of
hotspots (Fig.~\ref{fig:hr_image610}). The only 
feature with a significant surface brightness 
contrast with respect to the rest of the lobe is the 
shell--like edge visible in the outermost region of 
the eastern lobe (Fig.~\ref{fig:hr_image610}). The lobes 
are very symmetric in total flux density, size and 
spectral index (see Fig.~\ref{fig:cartoon} and 
Tab.~\ref{tab:source}). They are slightly 
asymmetric in the projected distance from the core: 
the centroid of the lobe West is at $\sim$65 kpc from the 
center, while that of the lobe East is at $\sim$ 50 
kpc. 

\section{Radio spectral analysis and physical parameters}\label{sec:spectra}

The radio observations presented in this paper, 
coupled with archival and literature data, allowed us 
to determine the integrated radio spectrum of 4C\,+24.36,
and to produce the image of the spectral index distribution
over the source (Sec.~\ref{sec:tot_sp}). This information
allowed us to derive the physical parameters of 4C\,+24.36
(Sec.~\ref{sec:parfis}). The analysis is carried out using 
the Synage++ package (Murgia 2000).

\subsection{Spectral analysis}\label{sec:tot_sp}

In order to derive the broad--band radio spectrum of 
4C\,+24.36, we used the databases CATS\footnote{http://cats.sao.ru} 
(Verkhodanov et al. 1997) and NED\footnote{http://nedwww.ipac.caltech.edu} 
to compile flux densities at different frequencies from the 
literature, and added our GMRT measurements (Tab.~\ref{tab:source}). 
The values from the literature are reported in Tab.~4. The flux 
densities marked with the symbol $\diamond$ in Tab.~4 were corrected 
for consistency with the absolute flux density scale of Baars et 
al. (1977) used in this paper and for the other measurements in 
Tab.~4. The correction factors were taken from Helmboldt et al. (2008).
The 74 MHz value was measured on the image from the VLA Low--frequency 
Sky Survey (VLSS\footnote{http://lwa.nrl.navy.mil$/$VLSS$/$}), 
and was corrected for the {\it clean bias} as described in
Cohen et al. (2007). The flux density at 327 MHz was measured
on the VLA--C array image of 4C\,+24.36, obtained from a new
reduction of public archival data (Obs.Id. AC598).

\begin{figure}
\centering
\plotone{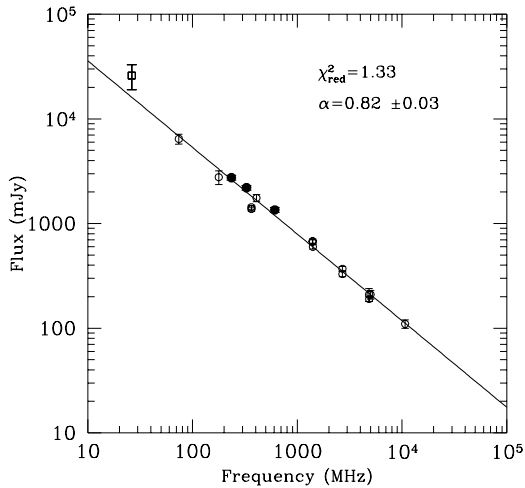}
\caption{Radio spectrum of 4C\,+24.36 between 26 MHz and 
10.7 GHz. The empty symbols are literature data (see 
Table~\ref{tab:table3}), while the filled circles are the 
GMRT 235, 327 and 610 MHz values (Tab. 4). The solid line 
is the power--law fit to the data. The uncertain 
26 MHz data point (empty square) was not included in the fit 
(see text). The value of $\alpha$ provided by the 
fit, along with the reduced $\chi^2$, are reported.}
\label{fig:spectrum}
\end{figure}

The source integrated spectrum between 26 MHz and 
10.7 GHz is shown in Fig.~\ref{fig:spectrum}. The GMRT 
data points align well with the data from the literature. 
We notice the extremely good agreement of the GMRT 327 MHz 
flux density with the measurement from the VLA image at the
same frequency (Tab.~4). The flux density at 26 MHz (empty 
square) was derived from observations with the Clarke Lake 
Radio Observatory with a resolution of $\sim$0.5 deg$^2$
(Viner \& Erickson 1975), and thus powerful nearby radio sources 
might contribute to the total flux density reported in Tab.~4. 
For this reason this value is very uncertain, and will not
be included in the following analysis of the radio spectrum.
\\
The spectrum in Fig.~\ref{fig:spectrum} can be described 
as a single power--law over the entire range 74 MHz--10.7 GHz, 
with a spectral index of $\alpha_{\rm obs}$=0.82. 
The solid line in the figure represents the fit with a 
simple power--law model, which provides $\alpha$=0.82$\pm$0.03.


\begin{table}\label{tab:flux2}
\caption[]{Literature radio data for 4C\,+24.36}
\begin{center}
\begin{tabular}{rcc}
\noalign{\smallskip}
\hline\noalign{\smallskip}
$\nu$     & Flux density & Reference\\
 (MHz)    &   (Jy)     &       \\
\hline\noalign{\smallskip}
26   & 26.00 $\pm$ 7.00 & (1) \\  
74   &  6.12 $\pm$ 0.47 & (2) \\
$\diamond$ 178 & 2.77 $\pm$ 0.42 & (3) \\
327  &  2.19 $\pm$ 0.11 & (4) \\
$\diamond$ 365 & 1.43 $\pm$ 0.07 & (5)\\  
$\diamond$ 408 & 1.76 $\pm$ 0.14 & (6) \\
1400 &  0.68 $\pm$ 0.03 & (7) \\
1400 &  0.67 $\pm$ 0.03 & (8) \\
1410 &  0.60 $\pm$ 0.04 & (9) \\
2700 &  0.33 $\pm$ 0.02 & (10) \\
2700 &  0.37 $\pm$ 0.02 & (9) \\
4860 &  0.19 $\pm$ 0.01 & (11) \\
4850 &  0.21 $\pm$ 0.03 & (12) \\ 
5000 &  0.20 $\pm$ 0.02 & (9) \\ 
10700&  0.11 $\pm$ 0.01 & (13)\\   
\hline\noalign{\smallskip}
\end{tabular}
\end{center}
\tablerefs{(1) Viner \& Erickson (1975); (2) VLSS image,
Cohen et al. (2007); (3) 4C; Pilkington \& Scott (1965), Gower et 
al. (1967); (4) VLA archival data (Obs. Id. AC598); (5) TXS, 
Douglas et al. (1996); (6) B2, Colla et al. 1970, Fanti et al. (1974); 
(7) FIRST, White et al. (1997); 
(8) NVSS, Condon et al. (1998); (9) PKS90, Wright \& Otrupcek (1990); 
(10) Bridle, Kesteven \& Brandie (1977); (11) GB6, Gregory et al. (1996); 
(12) Gregory \& Condon (1991); (13) Reich et al. (2000).}

\tablenotetext{}{The values marked with $\diamond$ were corrected for 
consistency with the Baars et al. (1997) scale.}
\end{table}


%
%

\begin{table}[h!]
\caption[]{Results of the point--to--point analysis}
\begin{center}
\begin{tabular}{ccccc}
\noalign{\smallskip}
\hline\noalign{\smallskip}
  & $\alpha_{inj}$ & $\nu_{\rm br}$ & t$_{\rm rad}$ & v$_{\rm growth}$ \\
  &            & (MHz)              &  ($10^{8}$) yr & (c)  \\
\hline\noalign{\bigskip}
jet+lobe West\tablenotemark{(b)} & 0.48$^{+0.07}_{-0.07}$ & 359$^{+37}_{-31}$ & 1.6  & 0.0017   \\
jet+lobe East \tablenotemark{(b)} & 0.45$^{+0.07}_{-0.07}$ & 316$^{+29}_{-26}$  & 1.7  & 0.0015   \\
&&&&\\
\hline\noalign{\smallskip}
\end{tabular}
\end{center}
\end{table}

%
%

\begin{figure}
\centering
\plotone{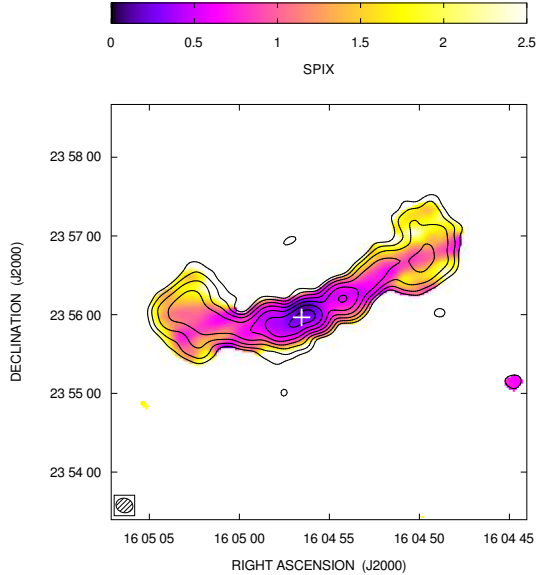}
\caption{Colour scale image of the spectral index 
distribution between 235 MHz and 610 MHz over 4C\,+24.36. 
The image has been computed from images with a restoring
beam of $13.0^{\prime \prime} \, \times \, 11.0^{\prime \prime}$. 
Overlaid are the 235 MHz contours at levels 3,6,12,24,48,96 and 
192 mJy b$^{-1}$. The white cross shows the position of the
radio core.}
\label{fig:spix}
\end{figure}

We obtained the image of the 235--610 MHz spectral index 
distribution over the source using two images produced 
with the same u--v range and same cellsize, restored with 
the same beam of $13.0^{\prime \prime} \, \times \, 
11.0^{\prime \prime}$, and corrected for the primary beam 
attenuation. The images were aligned, clipped at the 
3$\sigma$ level, and finally combined to obtain the spectral 
index image presented in Fig.~\ref{fig:spix}. The figure 
shows the presence of a flat region ($\alpha \sim 0.3 \pm0.1$) 
at the source center, corresponding to the inner $\sim$ 10 kpc 
portion of the jets (see Fig.~\ref{fig:cartoon}). On a larger scale, 
the spectral index gradually steepens along the jets up to 
$\alpha \sim 0.9 \pm 0.1$ where the jets merge into the lobes. 
Finally, a further steepening is observed in the lobe region where 
the spectral index has an average value $<\alpha>=1.5 \pm 0.2$.

%
%

%
%

\begin{figure}
\centering
\plotone{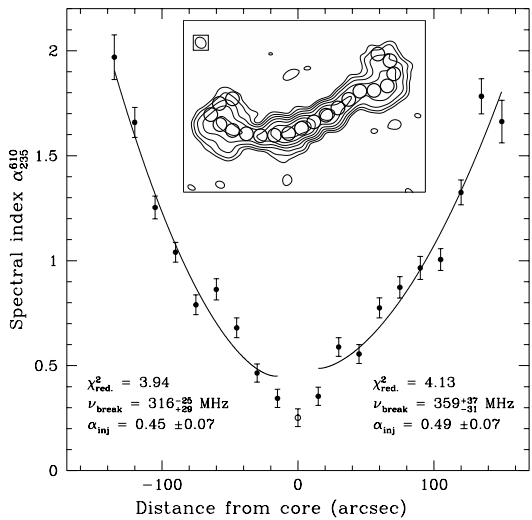}
\caption{Spectral index distribution, calculated
from the 235 and 610 MHz observations, along the 
source as a function of the distance from the core, derived 
using the circular regions shown in the insert. The radius 
of each region is 7.5$^{\prime \prime}$. The ellipse in the 
upper--left corner of the insert indicates the size of the 
radio beam (HPBW=$13.0^{\prime \prime} \, \times \, 
11.0^{\prime \prime}$). The contours correspond
to the 235 MHz image (same as in
Fig.~\ref{fig:spix}). The solid lines represent the best fit 
of the radiative model described in the text. The values of 
$\alpha_{\rm inj}$ and $\nu_{\rm br}$ provided by the fit, 
along with the reduced $\chi^2$, are also reported. The data 
point at $d\!=\!0$ (empty circle) was not used in the fit.}
\label{fig:steepening}
\end{figure}

Following Parma et al. (1999 and 2007; see also Giacintucci et 
al. 2007), we performed a fit of the observed spectral index trend 
along the source axis. We determined the average spectral index in 
7.5$^{\prime \prime}$ radius circular regions starting from the source
center and going outwards toward the lobes (see insert in 
Fig.~\ref{fig:steepening}). The size of the region was chosen to be larger 
than one beam, in order to sample independent regions. The derived 235--610 
MHz spectral index distribution as a function of the distance from the 
core is shown in Fig.~\ref{fig:steepening}. We notice that the two spectral 
trends are very similar once errors are taken into account.
%
%

%
%

We fitted these trends using a JP model (Jaffe \& Perola 1974), in 
which the timescale for continuous isotropization of the electrons 
is assumed to be much shorter than the radiative timescale. We also 
assumed that the break frequency $\nu_{\rm br} \, \propto 
\, d^{-2}$, where $d$ is the distance from the core. Such a relation 
is expected under the assumption that the expansion velocity of the 
source is constant (i.e. $d \, \propto \, t$), and reflects the fact 
that the age of the radio emitting electrons increases as one moves 
away from the nucleus, which is appropriate for FR I radio galaxies 
such as 4C\,+24.36. The best fit to the observed spectral trend is 
shown as solid line in Fig.~\ref{fig:steepening}. Note that the data 
point at d$=$0 (empty circle) was not used in the fit. The model 
yields similar values of the injection spectral index and break 
frequency for both regions, with $\alpha_{\rm inj} \sim 0.5$ 
and $\nu_{\rm break} \sim 360$ and $\sim$ 320 MHz for the western and
eastern parts, respectively (see also Tab.~5). 
 
\subsection{Physical parameters of 4C\,+24.36}\label{sec:parfis}

We assume that the relativistic particle and magnetic field 
energy densities are uniformly distributed over the source volume 
and in energy equipartition. The equipartition parameters 
are derived adopting a low--energy cut--off of $\rm \gamma_{min}$ 
in the energy distribution of the radiating electrons (where
$\gamma$ is the electron Lorentz factor), instead of the 10 
MHz--100 GHz frequency interval normally used in the standard 
equipartition equations (e.g. Pacholczyk 1970). This choice 
allows us to take into account the contribution from low energy 
electrons, since the assumption of a low frequency cut--off of 
10 MHz in the synchrotron spectrum tends to neglect the input from 
electrons with energy lower than $\gamma \sim$1500--500 for typical 
equipartition magnetic fields of $\sim$1--10 $\mu$G, respectively 
(Brunetti, Setti \& Comastri 1997). We adopt $\rm \gamma_{min}\!=\!10$ 
which corresponds to an energy cut--off of $\sim$5 MeV (Brunetti, Setti 
\& Comastri 1997; see also Parma et al. 2007). Furthermore, we use the 
flux density measurement at 235 MHz (Tab.~2), since the electron aging 
is less relevant at lower frequencies. A spectral index of 0.5, 
consistent with the values of the injection spectral index 
$\alpha_{\rm inj}$ derived in Sec.~\ref{sec:tot_sp} (Tab.~5), is adopted.
Finally, we assume a relativistic proton energy to electron energy ratio 
$\kappa$=1, and a filling factor $\phi$ of unity.  

%
%

\begin{table*}[t]
\label{tab:parfis}
\caption[]{Physical parameters of 4C\,+24.36 and its components}
\begin{center}
\begin{tabular}{ccccc}
\noalign{\smallskip}
\hline\noalign{\smallskip}
 & B$_{\rm eq}$ & P$_{\rm min}$ & U$_{\rm min}$ & u$_{\rm min}$ \\
 & ($\mu$G) & ($10^{-13}$ dyne cm$^{-2}$) & (10$^{57}$ erg) & ($10^{-12}$ erg cm$^{-3}$) \\
\hline\noalign{\bigskip}
total source & 5.00 & 7.8 & 3.7 & 2.3 \\
&&&&\\
jet+lobe West & 5.04 & 8.0 & 1.9 & 2.4 \\
jet+lobe East & 4.97 & 7.9 & 1.8 & 2.3  \\
&&&&\\
lobe West    & 3.14 & 3.1 & 0.9 & 0.9 \\
lobe East    & 3.17 & 3.1 & 0.9 & 0.9 \\
&&&&\\
jet West      & 8.99 & 25.0 & 0.7 & 7.5 \\
jet East     & 9.42 & 27.3 & 0.5 & 8.2 \\
\hline\noalign{\smallskip}
\end{tabular}
\end{center}
\end{table*}


The equipartition magnetic field B$_{\rm eq}$ computed in 
this way is listed in Tab.~6 for the entire source, and for
parts thereof. We found B$_{\rm eq} \sim 5$ $\mu$G both for 
the whole source and the western and eastern jet+lobe regions 
analysed in Sec.~\ref{sec:tot_sp}. A lower estimate is obtained 
for the lobes separately (B$_{\rm eq} \sim 3$ $\mu$G), while 
the highest values are found in the jets where B$_{\rm eq} 
\sim 9$ $\mu$G. We also derived the total energy U$_{\rm min}$, 
energy density u$_{\rm min}$, and minimum pressure 
p$_{\rm min}$ for the source and its components (Tab.~6).

The break frequency derived in Sec.~\ref{sec:tot_sp}
can be used to estimate the radiative age $\rm t_{\rm rad}$
of 4C\,+24.36. If the expansion losses can be neglected 
and the magnetic field is uniform across the source and 
constant over its lifetime, $\rm t_{\rm rad}$ (expressed in 
Myrs) is given by the following equation:

\begin{equation}\label{eq:trad}
{\rm t_{rad} = 1590\frac{B^{0.5}}{(B^2 + B_{CMB}^2)} [(1+z) \nu_{br}]^{-0.5}} 
\end{equation}

\noindent where B is the magnetic field intensity in $\mu$G, 
$\nu_{\rm br}$ is in GHz, and $\rm B_{CMB} (\mu G)= 3.2 \times 
(1+z)^2$ is the magnetic field strength with energy density 
equal to that of the CMB the redshift z. Adopting 
the equipartition magnetic fields (Tab.~6) in Eq.~\ref{eq:trad},
we estimate ${\rm t_{rad}} \sim 160$ Myr and $\sim$170 Myr for
the western and eastern parts of 4C\,+24.36, respectively (also
reported in Tab.~5). A growth velocity of $\rm v_{growth} 
\sim 0.002$c is found for both regions (Tab.~5). The latter quantity 
was derived as $\rm v_{growth} = LLS/t_{rad}$, i.e. assuming 
a constant velocity and using the largest linear size given 
in Tab.~2. It is worth noting the good agreement between the results 
obtained for two regions.

\section{Discussion of the radio results}\label{sec:disc}

\subsection{Jet orientation and velocity}\label{sec:theta}

The radio galaxy 4C\,+24.36 has WAT morphology in all the 
images presented in this paper (Figs.~\ref{fig:hr_image610} 
and \ref{fig:hr_images}), with two bright continuous jets 
and very extended radio lobes. Such components are very 
symmetric in total flux density and extent (Tab.~2), as
well as in spectral index properties and physical parameters 
(Tabs.~5 and 6). These results suggest that the major axis 
of 4C\,+24.36 most probably lies close to the plane of the sky.
\\
We searched for further hints of the source orientation
in the plane of the sky by means of a standard analysis based
on the jet asymmetry. Assuming that the jets are intrinsically 
symmetric with respect to the source core, the jet to counterjet 
brightness ratio $R$ can be used to constrain the geometry and 
velocity of the jets in the context of a relativistic beaming 
model (e.g. Lind \& Blandford 1985) according to

\begin{equation}\label{eq:ratio}
R=(1+\beta cos \theta)^{q}(1-\beta cos \theta)^{-q},
\end{equation}

\noindent where $\beta$ is the ratio of the jet velocity to 
the speed of light, $\theta$ is the orientation angle of the 
jet with respect to the line of sight, and $\alpha$ is the 
jet spectral index. If the jet emissivity is isotropic 
with no preferred direction for the magnetic field (Pearson 
\& Zensus 1987), the exponent $q$ is given by $2+\alpha$.

\begin{figure}
\centering
\plotone{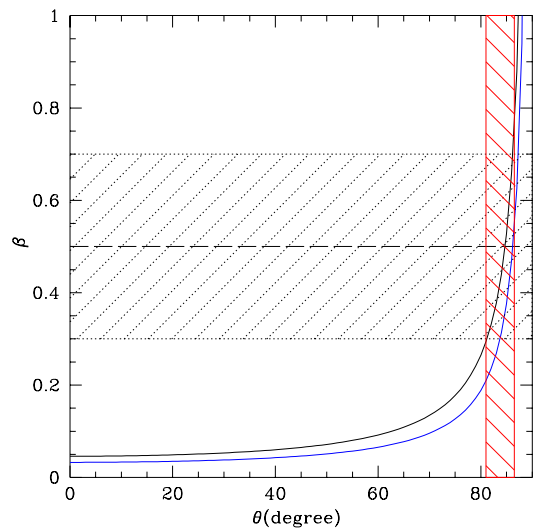}
\caption{Constraints on the angle $\theta$ between
the jet and the line of sight, and the intrinsic
jet velocity $\beta$ for 4C\,+24.36. The solid black 
line is the curve obtained from the jet--counterjet 
ratio in the isotropic case (i.e., using $q=2+\alpha$ 
in Eq.~\ref{eq:ratio}), while the solid blue curve is 
given by $q=3+2\alpha$. The long-dashed line, corresponding 
to $\beta=0.5$, is an average value for the jet speed 
in WATs (the black dashed region represents the 
1$\sigma$=0.2 dispersion). The red dashed region shows the 
allowed values for $\theta$ in 4C\,+24.36 in the isotropic
case, corresponding to $\beta=0.5\pm0.2$.}
\label{fig:beta}
\end{figure}

Using the 0.7$^{\prime \prime} \times 0.5^{\prime \prime}$ 
resolution image at 4.9~GHz (see insert in Fig.~\ref{fig:cartoon}), 
and assuming the jet East to be the counterjet, we estimated
$R$ within the inner $\sim 5$ kpc from the core. Assuming 
$\alpha=0.5$ for the jet emission, we found $R=1.3$ which 
yields $\beta \cos \theta$=0.05 (Eq.~\ref{eq:ratio}). 
Fig.~\ref{fig:beta} shows the corresponding constraints 
on $\theta$ and the intrinsic jet velocity $\beta$ (solid 
black curve). Jetha, Hardcastle \& Sakelliou (2006) found that the average value 
for the inner jet speed in a relatively large sample of WAT radio 
galaxies is $\beta=0.5\pm0.2$. This velocity is in general agreement 
with the values found for FRI radio galaxies, whose properties are 
consistent with the hypothesis that jets slow down from relativistic 
to sub--relativistic velocities on scales of 1--10 kpc (Parma et al. 
1994). Adopting $\beta=0.5\pm0.2$ for 4C\,+24.36, the black solid 
curve in Fig.~\ref{fig:beta} shows that the allowed jet orientation 
angle is in the range $\theta \sim 81^{\circ} $-- $87^{\circ}$ 
(red dashed region). This suggests that the source probably lies 
at a very small angle with respect to the plane of the sky.
\\
If we use the exponent $q=3+2\alpha$ in Eq.~\ref{eq:ratio}, which
is valid in the case of a perfectly ordered magnetic field
parallel to the jet axis (Begelman 1993), we obtain very similar 
constraints for the jet orientation, as shown by the solid blue 
curve in Fig.~\ref{fig:beta}. In this case $\theta \sim 
84^{\circ}$--$88^{\circ}$.

%
%

%
%

\subsection{Morphological distortions}\label{sec:morph}

The radio structure of 4C\,+24.36 is characterised 
by clear distortions occurring over different linear 
scales. One of most prominent features is the number 
of sharp wiggles which the jets undergo within the 
central $\sim 25$ kpc (projected) from the core 
(Fig.~\ref{fig:profile}). On the large scale, the 
overall structure of the source appears slightly 
bent toward the North in a C--shape which is typical 
of WAT sources (Fig.~\ref{fig:hr_image610} and 
Fig.~\ref{fig:hr_images}). This suggests that 
different physical processes may play a role in the 
origin of the morphological distortions observed on the 
different scales.

\subsubsection{Jet wiggles}\label{sec:wiggles}

Jet oscillations have been observed in other FRI 
radio sources of similar luminosity as 4C\,+24.36, 
e.g. the well studied radio galaxies 
3C\,31 (e.g., Blandford \& Icke 1978; Andernach et 
al. 1992; Laing \& Bridle 2002) and 3C\,449 (e.g., Andernach et al. 1992; 
Feretti et al. 1999). Both sources are characterized 
by a very large total extent (LLS $\sim$ 800 kpc and 
600 kpc, respectively) which places them among the 
class of giant radio galaxies. They are associated 
with the dominant members of two nearby poor groups 
of galaxies (z=0.17), and both galaxies are likely 
undergoing a gravitational encounter with a close 
bright companion.

%
%

\begin{figure*}[t]
\centering
\epsscale{0.75}
\plotone{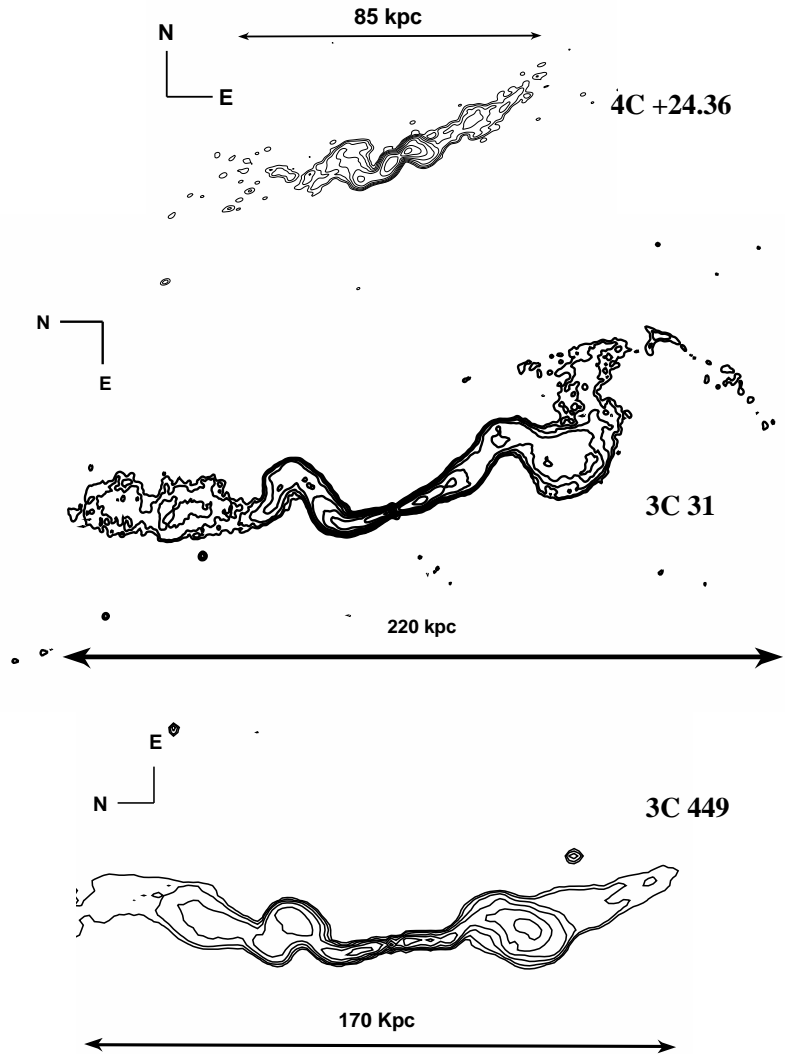}
\caption{Comparison of the radio morphology of
4C\,+24.36, 3C\,31 and 3C\,449 (note that the last
one has been reversed in order to better highlight the 
similarities). {\it Upper panel} -- 1.4 GHz image of 4C\,+24.36 
obtained from the combination of VLA A and B--Array observations 
from the public archive (Obs.Id. 074). The HPBW is 
3.6$^{\prime \prime} \times 1.8^{\prime \prime}$, 
p.a. -65$^{\circ}$. The first contour is 0.15 
mJy b$^{-1}$. {\it Central panel} -- VLA--BnA 1.4 
GHz image of 3C\,31 from the public archive (Obs.Id.
AF263). The HPBW is 4.1$^{\prime \prime} \times 3.7^{\prime \prime}$, 
p.a. -65$^{\circ}$. The first contour is 0.25 mJy b$^{-1}$.
{\it Lower panel} -- VLA--B 1.4 GHz image of
3C\,449 from the public archive (Obs.Id. AK319). The 
HPBW is 4.5$^{\prime \prime} \times 3.9^{\prime \prime}$, 
p.a. 86$^{\circ}$. The first contour is 0.35 mJy b$^{-1}$.}
\label{fig:morph}
\end{figure*}

%

In Fig.~\ref{fig:morph} we compare the radio morphology 
of 4C\,+24.36 (upper panel) to the structure of 
the well-known radiogalaxies 3C\,31 
and 3C\,449 (central and lower panels, respectively). 
All images were obtained from a new reduction of 1.4 GHz 
observations  from the VLA public archive, and have a resolution 
of $\sim 4^{\prime \prime}$ (see Figure caption for details). 
Note that the images of 3C\,31 and 3C\,449 have been rotated
and reversed in order to highlight the morphological 
similarities. We also point out that the image of 3C\,449 
shows only the central region of the source, whose total 
extent at this resolution is $\sim 20^{\prime}$ (i.e. $\sim$ 
420 kpc; see Feretti et al. 1999). Despite the different 
linear size, the similarity of the pattern of wiggles in 
the three sources is striking. However we can notice some 
important differences: {\it i)} the wiggles in 4C\,+24.36 
occur over a smaller projected distance from the nucleus 
(i.e. between $\sim$7 kpc and $\sim$25 kpc; see Fig.~\ref{fig:profile}) 
than in 3C\,31 and 3C\,449, where the first bend is observed 
at $\sim 20$ kpc; {\it ii)} the inner jets in 4C\,+24.36 
show a much stronger deviation from linearity than 3C\,31 
and 3C\,449. In particular, the western jet appears 
already bent at $\sim$3 kpc only from the core (see insert 
in Fig.~\ref{fig:cartoon}).
\\
\\
Jet oscillations can be ascribed to various reasons, which 
include {\it a)} orbital motion, {\it b)} precession 
of the nuclear collimator, and {\it c)} jet instabilities, 
such as Kelvin-Helmholtz (K--H) instabilities, at 
the jet boundary. 

\begin{itemize}

\item [{\it a)}] The orbital motion is expected to generate 
a single oscillation wavelength with fixed amplitude and 
mirror symmetry of the jet with respect to the core. Given the 
presence of a close bright companion, it has been 
suggested that the orbital motion might indeed be responsible for 
the jet wiggles in 3C\,31 (Blandford 
\& Icke 1978) and 3C\,449 (Perley, Willis \& Scott 1979). 
However no bright companion is visible on the available 
optical images of 4C\,+24.36 (e.g. Fig.~\ref{fig:hr_image610}), 
thus it seems difficult to ascribe the jet oscillations to 
the presence of an orbiting system. Furthermore, with plausible 
galaxy masses and the orbital separation required to produce 
the observed lateral displacement of the jets, the orbital 
period would be far larger than the jet age allows.

\item [{\it b})] Precessional motion is expected to lead to a 
single oscillation wavelength with amplitude linearly growing 
with the distance from the nucleus and anti--symmetric pattern. 
Given the approximate mirror symmetry of the initial jet 
oscillations in 4C\,+24.36, we can reasonably rule out the 
precessional model. 

\item [{\it c})] Confined radio jets are known to be unstable 
against K--H instabilities at their boundary (Hardee 1987). The 
jet oscillations can be then interpreted as the result of helical 
motion arising from small perturbations at the jet origin, 
amplified by growing K--H instabilities (Hardee 2003; Savolainen 
et al. 2006). These instabilities can be triggered for example 
by random perturbations of the jet flow (e.g., by jet--cloud 
interactions; G\'omez et al. 2000), or by variation in the jet 
injection direction from the nucleus (e.g., due to the 
orbital motion of a supermassive binary black hole in the host 
galaxy). It has been shown that the K--H helical distortion 
waves propagating along the jet are able to displace the whole 
jet and produce large scale helical structures (Hardee 1987). 
Thus K--H instabilities are an appealing possibility to interpret 
the wiggles observed in 4C\,+24.36. Unfortunately the resolution of 
our present radio data is not high enough to allow us to 
determine the characteristics of a possible K--H wave propagating 
along the jet. 
\end{itemize}

Further information on the fine brightness structure and 
polarization properties of the jets from higher resolution 
radio observations would be required to shed light on the 
possible mechanisms responsible for the oscillations. 
Moreover, forthcoming high resolution X--ray {\it Chandra} 
observations (Cycle~9) will allow us to carry out a 
detailed structural comparison of the radio and X--ray 
emission in the jet regions. 

\subsubsection{Large scale bending}\label{sec:bending}

The investigation of the physical mechanism responsible for the 
wide--angle tail formation poses a number of interesting problems 
which are not completely understood yet. The earliest models developed 
for the origin of WATs invoked ram pressure resulting 
from the motion of 
the associated galaxy through the surrounding intracluster medium 
(e.g., Owen \& Rudnick 1976; Begelman, Rees \& Blandford 1979). 
However, WATs are usually associated with the central dominant galaxies 
in groups and clusters, and they are not expected to have large velocities 
relative to the ICM (Burns 1981), as they usually reside at the center 
of the cluster gravitational potential well (Merritt 1984; Bird 1994). 
Thus, ram pressure alone seems unable to explain the bends observed 
in most of the WATs (e.g., Eilek et al. 1984; O'Donoghue, Eilek \& Owen 
1993).
\\
One possibility is that if the radio jets are less dense than the
external gas, buoyancy forces might contribute to the bending of the tails, 
as they drag the radio plasma towards lower density regions in the ICM 
(e.g., Burns \& Balonek 1982). In real situations, both ram pressure by
the galaxy motion and buoyancy are expected to play a role, with buoyancy 
forces dominant at larger radii (Sakelliou et al. 1996). An alternative 
scenario is that the large scale bulk motions in the ICM, resulting from
cluster mergers, may provide the necessary ram pressure to distort the 
structure of the central dominant radio galaxy (e.g., Pinkney, Burns 
\& Hill 1994; G\'omez et al. 1997). Large scale X--ray substructure
and merging signatures have been indeed observed in many clusters hosting a 
WAT (Burns et al. 1994). Finally, an appealing possibility to explain 
the existence of WATs at the center of non--merging clusters is the possible 
connection of the radio source bending with the process of gas sloshing in 
the cluster core thought to be responsible for the formation of cold fronts 
(Ascasibar \& Markevitch 2006). For example, such a connection might be 
present in A\,2029, a very relaxed cluster hosting a WAT and a pair of cold 
fronts at its center (Clarke, Blanton \& Sarazin 2004; Ascasibar \& 
Markevitch 2006 ).
\\
\\
The optical and X--ray properties of AWM\,4 (Koranyi \& Geller 2002 and 
OS05, respectively) indicate that the cluster has not 
recently experienced major episodes of merging (see also Sec.~\ref{sec:environment}). 
Moreover, the XMM--Newton images do not show any sharp surface brightness 
edges suggestive of the existence of cold fronts in the core (OS05). For 
these reasons we interpret the bending in 4C\,+24.36 as a result of the
interplay of ram pressure (driven by the motion of NGC\,6051) and buoyancy 
forces. Following Sakelliou et al. (1996), we can derive a constraint on 
the galaxy velocity relative to the ICM, as a function of the ratio of the 
jet density to the ICM density, at the point where the components of ram 
pressure and buoyancy in the direction orthogonal to the jet balance
each other (the {\it turnover point}). At smaller distances than this point, the jet 
bending is expected to be mostly determined by ram pressure, while 
buoyancy forces dominate at larger distances. Furthermore, the bending 
does not depend on the jet velocity at the turnover point (Sakelliou et 
al. 1996). 
\\
We assume that the radio jets can be described as a steady 
flow of a non--relativistic plasma (Bridle \& Perley 1984), and adopt
a coordinate system in which the axes always lie parallel and perpendicular 
to the jet. At the {\it turnover} point the component of the galaxy 
velocity in the plane of the sky ($\rm v_{gal}$) can be expressed as 
(see Sakelliou et al. 1996, and Smol\v{c}i\'{c} et al. 2007 for details),
\begin{equation}\label{eq:vgal}
\rm v_{gal}^2 = \frac{ 3\beta k T_{gas} h }{ \mu m_p r_c }
            \frac { { r_{to}/r_c} }{ 1 +  \left ({ r_{to} / r_c }\right ) ^2}
            \frac{\bf \hat{r}\cdot \hat{n}}{{\bf \hat{v}_{gal}\cdot \hat{n}}}
            \left( 1 - \frac{\rho_{jet}}{\rho_{gas}}\right),
\end{equation}
\noindent 
where $\rm \bf{\hat{n}}$ is the normal vector to the jet in the plane 
of the sky, $\rm \bf{\hat{v_{gal}}}$ is the unity vector in the direction of 
the galaxy velocity, $\rm \bf{\hat{r}}$ is the unity vector in the direction from 
the cluster center, $\beta$ and $\rm r_c$ are the slope and core radius of the 
X--ray surface brightness when fit to a hydrostatic $\beta$ model, kT is the 
ICM temperature, $\rm r_{to}$ is the projected distance of the turnover point 
from the cluster center, h is the width of the jet at this point (corrected for 
the resolution of the telescope), $\mu$ is the mean molecular weight ($\mu$=0.6), 
$\rm m_p$ is the proton mass, and $\rm \rho_{gas}$ are the density of the jet 
and ICM.

\begin{figure}
\centering
\epsscale{1.00}
\plotone{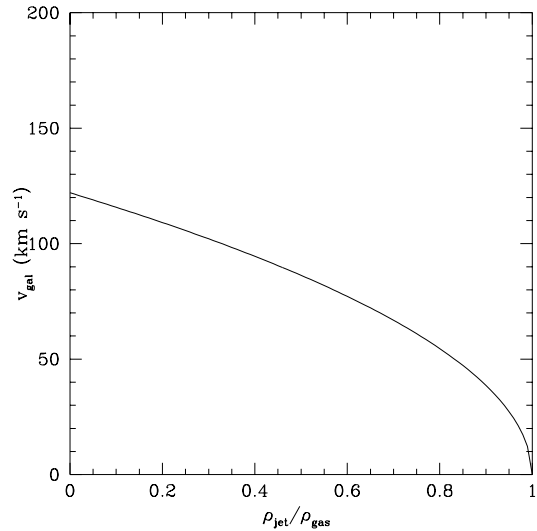}
\caption{The transverse component of the 
velocity of NGC\,6051 (in the plane of the sky), as a function 
of the ratio of the jet density to external gas density.}
\label{fig:vgal}
\end{figure}

A close examination of Figs.~\ref{fig:hr_image610} and \ref{fig:profile} 
suggests that the last change in direction of the jets (points E4 and W4) 
might be identified as the turnover point for 4C\,+24.36.
In this case $\rm r_{to} \sim 25$ kpc. The average gas temperature 
in AWM\,4 is kT=2.5 keV (Tab.~1), and the fit of the radial profile 
of the cluster X--ray surface brightness with a single $\beta$ model 
provides $\rm r_c=67$ kpc and $\beta=0.67$ (OS05). Thus, if $h\!=\!8$ kpc, 
as estimated from Fig.~\ref{fig:hr_image610}, then from Eq.~\ref{eq:vgal} 
we can derive the galaxy velocity as a function of $\rm \rho_{jet} / 
\rho_{gas}$ (Fig.~\ref{fig:vgal}). In the limit $\rm \rho_{jet} / 
\rho_{gas} \rightarrow 0$, we obtain an upper limit for the NGC\,6051 
velocity relative to the ICM of $\rm v_{gal} \ltsim 120$ km s$^{-1}$. 
\\
We stress that $\rm v_{gal}$ is the velocity component in the plane 
of the sky, therefore the derived value is a lower limit on the 
total speed of NGC\,6051. On the other hand, the comparison of the NGC\,6051 
redshift to the average redshift of the other cluster members provides 
an estimate of the line of sight component of its velocity. Koranyi \& 
Geller (2002) report a recession velocity of 9703$\pm$46 km s$^{-1}$ 
for NGC\,6051, and an average velocity for the cluster members within 
a radius of 0.5 $h^{-1}$ Mpc of 9722$\pm$98 km s$^{-1}$. 
Thus NGC\,6051 seems to be relatively at rest with respect to the 
potential well of AWM\,4. This is consistent with the fact that the 
galaxy position is coincident with the peak of the cluster X-‐ray 
emission (Koranyi \& Geller 2002; OS05), and is also in agreement 
with the velocities usually found for the dominant galaxies in 
non--merging groups and poor clusters of galaxies (e.g., Beers et 
al. 1995).


\subsection{Source age}\label{sec:age}

Our spectral analysis led us to an estimate of the synchrotron age 
of 4C\,+24.36 of $t_{\rm rad} \sim 160$ Myr (Tab.~5). This age is 
significantly higher than the typical values derived 
(using a similar approach to the one adopted in Sec.~\ref{sec:parfis}) 
for currently active radio galaxies of similar luminosity, as for 
example the sources in the B2 sample which have a median age of 16 
Myr (Parma et al. 1999, 2007). The age of 4C\,+24.36 is indeed more 
similar to the values derived by Parma et al. (2007) for a sample of 
candidate dying radio galaxies, which have a median age of 63 Myr, 
with values up to 170 Myr. 
\\
 Giacintucci et al. (2007) have recently analysed a sample 
of cD radio galaxies in rich and poor clusters of galaxies, 
and found radiative ages $ > 10^{8}$ years for 4 of the 7 sources 
with available estimates of the age. They argued that one of them 
(A\,2372) is a restarted radio galaxy, where both the new activity 
and the relic radio plasma from the previous AGN burst are 
visible. Another two sources (A\,2622 and MKW\,3s) are most likely dying 
radio galaxies, where the nuclear engine has switched off and a relic 
phase is currently ongoing. However, the total spectral index of 
4C\,+24.36 ($\alpha \sim 0.8$ in the 74 MHz--10.7 GHz range) is 
typical of a currently active radio galaxy rather than a relic source, 
where $\alpha$ can be much steeper ($\gtsim$2; e.g. Parma et al. 2007; 
Giacintucci et al. 2007). Furthermore, 4C\,+24.36 clearly shows active 
nuclear emission at high frequencies 
(see insert in Fig.~\ref{fig:hr_image610}). 
Thus it seems rather unlikely that we are dealing with a dying radio 
galaxy. 

We would like to point out here that the estimate of the radiative age 
in Tab.~5 is based on the assumption  
that the energy losses due to the source expansion can be neglected, 
and that the spectral steepening is due to electron aging in a magnetic 
field which is constant and uniform across the source. 
The role of a decreasing magnetic field along the source structure 
has been considered in alternative models for classical double radio
galaxies (e.g., Blundell \& Rawlings 2000, and references therein),
where the magnetic field decrease is reflected in a steepening of
the spectral index, with no direct implication for the source aging.

A detailed calculation of the evolution of the electron energy 
spectrum under the combined effect of radiative losses and adiabatic 
expansion is difficult, and it is beyond the purpose of the present 
work. However, we can derive a crude estimate of the effects of 
possible expansion losses in 4C\,+24.36 by considering that the cross 
section of the radio plasma increases by a factor of $\sim$2 in the 
transition from the jets to the lobes (Fig.~\ref{fig:hr_image610}; Tab.~3). 
If we further consider that the Inverse Compton losses are as important 
as synchrotron losses ($\rm B \sim B_{CMB} = 3.4$ $\mu$G; Tab.~5), we may 
expect that the break energy could be lower by an extra factor of $\sim$ 0.6 due 
to the expansion of the source. Moreover, because of the conservation of 
the magnetic field flux, we would expect a decrease of the magnetic field 
strength by a factor of $\sim$ 0.25. The combination of these effects 
would lead to a break frequency which is about one order of magnitude lower 
with respect to the case in which there is no expansion of the source. 
As a result, given that $\rm t_{rad} \propto \nu_{br}^{-0.5}$ 
(Eq.~\ref{eq:trad}), the age of the source would artificially appear a 
factor of $\sim$ 3 higher than its real value. This implies that in the
case of 4C\,+24.36, the real age would be of the order of $\sim$ 50 Myr.
Thus the age reported in Tab.~5 should be considered an upper limit to 
the real age of the source, and consequently the computed velocity 
growth as a lower limit.

\section{Gas heating at the core of AWM\,4?}\label{sec:xray}

The XMM--Newton observation of the core of AWM\,4 (Obs. ID. 0093060401; 
OS05) reveals that the ICM is isothermal at about 2.5~keV out 
to at least 160 kpc from the center, even though the cooling 
time in the middle is about 2~Gyr. In such systems, to account for 
the absence of the establishment of a cool core, many possible heating 
mechanisms have been proposed, including gravitational heating due to 
orbital motion and dynamical friction (e.g., Dekel \& Birnboim 2008), 
galaxy and sub-cluster mergers (e.g., G{\'o}mez et al. 2002), 
supernovae (e.g., Efstathiou 2000), turbulent mixing in the ICM 
(e.g., Kim \& Narayan 2003a), thermal conduction from the outer parts 
of the cluster (e.g., Kim \& Narayan 2003b), and conduction of energy 
from cosmic rays (e.g., Guo \& Oh 2008). However, the most promising 
energy injection models invoke the presence of a central AGN, and various 
modes of transport of the output energy from the supermassive black hole 
to the ICM have been considered (e.g., Binney \& Tabor 1995; Roychowdhury et 
al. 2004; Mathews et al. 2004; Nusser et al. 2006; Jetha et al. 2008). 
Here we conclude the detailed study of the remarkable radio galaxy 4C\,+24.36,
in the core of the relaxed yet isothermal cluster AWM\,4, by assessing
the role of the AGN in the prevention of the establishment of a 
cool core in this system.

\subsection{4C\,+24.36 and the cluster environment}\label{sec:environment}

\begin{figure}
\centering
\plotone{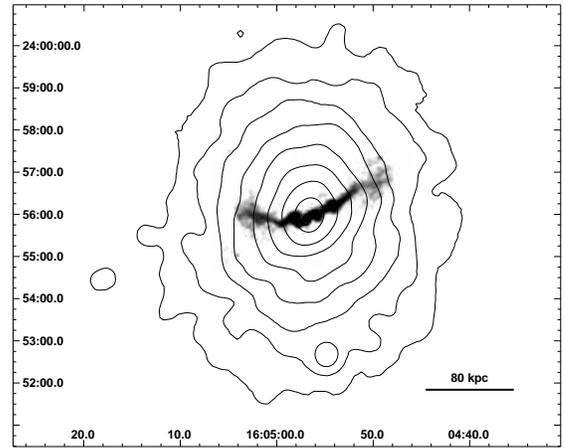}
\caption{Overlay of the grey--scale radio image
at 610 MHz of 4C\,+24.36 (same as Fig.~\ref{fig:hr_image610})
on the X--ray contours of AWM\,4 from the XMM--Newton
adaptively smoothed image, obtained by combining the 
observations with 
the PN and MOS cameras. The X--ray contour levels
are logarithmically spaced by a factor $\sqrt{2}$ .} 
\label{fig:xmm}
\end{figure}

The overall optical properties of AWM\,4 have been investigated by 
Koranyi \& Geller (2002). It is a relatively poor cluster containing 
about 30 members centered on NGC\,6051, within a projected radius of 
$0.5\,h^{-1}$ Mpc, with a velocity dispersion of $\sigma_{\rm} = 439$ 
km s$^{-1}$ (Tab.~1). Morphological segregation is clearly evident
within the cluster galaxy population, with the early--type galaxies more 
concentrated toward the core region (i.e., around NGC\,6051) than the 
late--type galaxies. No substructure has been detected in the projected 
galaxy distribution and/or in the velocity space, suggesting that AWM\,4 
is a relatively relaxed cluster, which has not recently experienced major 
mergers or infall of substantial subclusters.
\\
The relaxed nature of this system is further supported by the 
analysis of the XMM--Newton observations of the cluster by OS05 
(see also the recent study by Gastaldello et al. 2008). In 
Fig.~\ref{fig:xmm}, we show the XMM--Newton image of AWM\,4 
(contours, adaptively smoothed from a combination of the MOS and PN 
data), on which the GMRT 610 MHz image of 4C\,+24.36 (grey scale) has 
been superposed. The cluster X--ray surface brightness shows a very 
regular and relaxed appearance free of any visible substructures, 
with a steep increase of the brightness toward the center. The radio 
source extends over the whole core region of the cluster (core radius 
$\sim 70$ kpc as provided by the fit of the X--ray surface brightness 
to a single hydrostatic $\beta$ model; OS05). No significant feature 
in the hot gas distribution seems associated with the region filled by 
the radio emission or surrounding it. In particular, at the location 
of the radio lobes, we do not observe any depression in the X--ray 
surface brightness, which might indicate the presence of cavities in 
the ICM, evacuated by the radio galaxy, as is often observed in many 
relaxed clusters and groups of galaxies (see for instance the reviews 
by Blanton 2004, McNamara \& Nulsen 2007, and references therein).


\subsection{Pressure balance}


\begin{figure}
\centering
\plotone{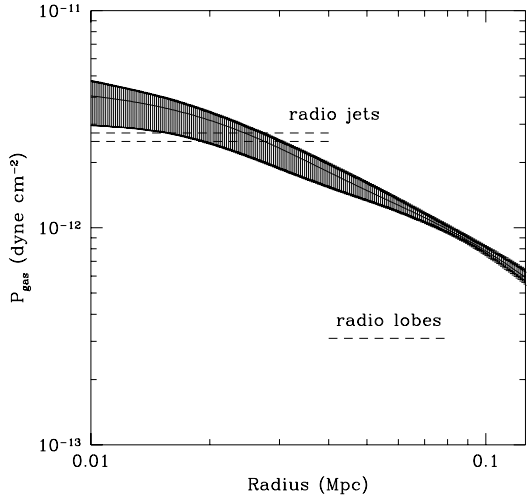}
\caption{Radial profile of the thermal gas pressure 
in the central region of AWM\,4 from the XMM--Newton 
data (O'Sullivan et al. 2005). The dashed error 
region is for 1$\sigma$ uncertainties. Dashed lines 
indicate the minimum synchrotron pressure in the radio 
jets and lobes of 4C\,+24.36.}
\label{fig:pressure}
\end{figure}

Fig.~\ref{fig:pressure} shows the profile of the thermal 
gas pressure ($\rm P_{gas}$) as a function of the distance
from the cluster center, inferred from the best fitting
X--ray temperature and density models as part of the mass
analysis described in OS05. In this plot, we also show the 
minimum pressure $\rm P_{min}$ of the radio 
jets and lobes of 4C\,+24.36 (dashed lines), estimated assuming 
equipartition arguments (Sec.~\ref{sec:parfis}; Tab.~6). From 
Fig.~\ref{fig:pressure} it is clear that the jet minimum 
pressure is comparable to $\rm P_{gas}$ over most of the jet 
length. On the contrary, the radio lobes appear underpressured 
with respect to the external medium by a factor ranging 
from $\sim$3 to $\sim$5. We point out that the assumption
of standard equipartition equations, i.e., without the 
low--energy cut--off of $\rm \gamma_{min}$ in the electron energy 
distribution (Sec.~\ref{sec:parfis}), would lead to 
a higher pressure imbalance in the lobe regions of the 
order of $\rm P_{gas}/P_{min} \sim 6-12$. We also notice 
that projection effects are not expected to play a 
significant role in the comparison in Fig.~\ref{fig:pressure}, 
as the source is approximately in the plane of the sky 
(Sec.~\ref{sec:theta}). 

A similar lack of pressure balance has been found in the large 
scale components of other FRI radio sources, where, based 
on standard equipartition arguments, the ratio $\rm P_{th}/P_{min}$ 
is usually $>1$, with values up to $\sim$100 (e.g., Feretti,
Perola \& Fanti 1992). This suggests a departure from the 
minimum energy condition within the radio lobes. In particular,
 
\begin{itemize}

\item[{\it i)}] the lobes could be far from equipartition, 
with a higher relativistic particle or magnetic field energy 
density.

\item[{\it ii)}] Assuming that equipartition still holds, 
the lobes may contain an energetically dominant population of 
non--radiating relativistic particles (protons) which contribute 
to the internal pressure. The minimum pressure is a weak function 
of the energy ratio $\kappa$ between relativistic protons and 
electrons, i.e., P$_{\rm min} \propto (1+\kappa)^{4/7}$. 
The values of P$_{\rm min}$ in Tab.~6 were estimated 
assuming $\kappa=1$. Thus, in order to achieve pressure balance 
in the lobe regions of 4C\,+24.36, we should require a 
contribution of relativistic protons such as $\kappa \sim$ 6--22, 
which is in agreement with the estimates for other radio 
sources of similar luminosity (e.g. Feretti, Perola \& Fanti 1992).

\item[{\it iii)}] The balance could also be achieved (in the equipartition 
hypothesis) assuming that the plasma has a filling factor 
$\phi \sim 0.05$--$0.2$ (according to P$_{\rm min} \propto 
\phi^{-4/7}$), instead of unity. Filamentary structures 
have been detected in high resolution radio images of a number of 
radio sources which may indeed indicate filling factors smaller 
than 1 (e.g., Fornax A; Fomalont et al. 1989; M87, Hines, Owen \& Eilek 1989).

\item[{\it iv)}] An additional pressure component might be 
provided by the presence of thermal plasma within the lobes. The
jets could accrete material as they move through the external
medium (e.g., Bicknell 1994) and transport it into the lobes.
This would be in line with the fact that the pressure imbalance
increases with the distance from the nucleus (Fig.~\ref{fig:pressure}). 

\end{itemize} 

\subsection{Energy output from 4C\,+24.36}

The analysis of the {\it XMM-Newton} observations in OS05 
revealed that AWM\,4 occupies an unusual place among groups and clusters (see 
also Gastaldello et al. 2008). The combination of the regular X--ray 
emission and monotonic increase of brightness toward the center would 
usually suggest a relaxed cluster with a well--established cool core. 
Such an expectation would be further supported by the abundance profile 
which shows a clear metallicity decline going from the cluster center 
to larger radii (OS05; Gastaldello et al. 2008). However, the X--ray 
spectral fitting reveals that AWM\,4 does not possess a cool core. 
The azimuthally averaged temperature profile is approximately constant 
out to a distance of at least $\sim$160 kpc from the center, without 
any sign of the substantial temperature drop in the center expected in 
an ordinary cool core cluster (OS05; Gastaldello et al. 2008).

Given the absence of any significant signature of a recent merger 
event which may have disrupted the cool core in AWM\,4 
(Sec.~\ref{sec:environment}), OS05 suggested that the AGN in the 
central galaxy NGC\,6051 has been active for enough time to (re--)heat 
the surrounding gas in the core. Assuming that a cool core existed 
in the cluster before the re--heating process, OS05 estimated 
that the total energy required to produce the observed isothermal 
profile is $\sim 9 \times 10^{58}$ erg. 

Using a spectral index $\alpha=0.5$ (consistent with the injection
spectral index provided by the spectral fit; Tab.~5), we calculate 
the total radio luminosity of 4C\,+24.36 over the frequency range 
10 MHz--100 GHz to be L$_{\rm radio} \sim 5.8 \times 10^{41}$ 
erg s$^{-1}$. In Sec.~\ref{sec:parfis} we estimate a synchrotron 
lifetime of the radiating electrons in the source of $\rm t_{rad}
\sim$ 160 Myr (Tab.~5). This value implies that the energy deposited
in the ICM by the source over its lifetime is E$_{\rm tot,\, radio}
\sim 2.9 \times 10^{57}$ erg. Thus the radio source seems to be at
least a factor of $\sim$30 too underluminous to balance the cooling in
this system. Indeed, in the sample of clusters studied by B\^irzan et
al. (2004, see also McNamara et al. 2007), the median ratio of the jet
power to the synchrotron power is  $\epsilon\sim 120$, the mean
value $\langle \epsilon \rangle\sim 4700$ being dominated by a few
extreme cases. In the case of AWM4, we should point out that this ratio 
could be much higher than 30, if we consider that the age of the source
might be up to $\sim$ 3 times less than the value of 160 Myr quoted
above, as we have discussed in Sec.~\ref{sec:age}.

On the other hand, the total energy output of a radio source can be 
considerably higher than what estimated from its synchrotron luminosity, 
since the total radio power is thought to be dominated by the mechanical 
work done by the jets on the ICM during the source expansion (e.g. Scheuer 1974). 
One mode of doing so is by inflating bubbles filled with a relativistic fluid
in the ICM, which would expand and rise, imparting energy to the ICM 
(Churazov et al. 2002,  Nulsen et al. 2003, Nusser et al. 2006, McNamara 
\& Nulsen 2007). This is supported by the direct detection of significant 
cavities in the X--ray surface brightness of several clusters, which might 
represent these rising bubbles (Fabian et al. 2000, 
B\^irzan et al. 2004 \& 2006, Allen et al. 2006, Gitti, Feretti \& Schindler 2006,
Jetha et al. 2008).

The enthalpy H of a rising adiabatic bubble is the sum of the $pV$ work
done to displace the X--ray emitting gas, and its thermal energy, and 
is given by $\frac{\gamma}{\gamma - 1} pV$, where $p$ is the pressure of
the surrounding ICM, $V$ the volume of the bubble and $\gamma$
the ratio of specific heats of the gas. For an ideal
relativistic gas, 
$\gamma\!=\! 4/3$, and $H=4pV$. In case of a non-relativistic
plasma filling the bubble, and also in the presence of significant
magnetic fields in the lobes, the value of enthalpy
could be a factor of 2 lower. Churazov
et al. (2002) argue that almost all of this enthalpy can be
transferred as thermal energy to the ICM. 

To assemble the equivalent of $\sim 9 \times 10^{58}$~erg of mechanical
energy, which is needed to produce the isothermal profile in the core of
AWM\,4, as estimated by OS05, with the lobe pressure being $p=3.1\times
10^{-13}$ dyne~cm$^{-2}$ (see Table~6), would require the entire
enthalpy ($=4pV$) of a spherical bubble of radius of over $\sim$80
kpc. Inflating this bubble against a pressure as high as jet
pressure, which is an order of magnitude higher, would require a bubble
of half this size, whereas a value of enthalpy that is lower than
$4pV$ would only require a larger bubble. Such an extensive
displacement of plasma in the ICM would be easily apparent in the
{\it XMM--Newton} observation of AWM\,4, where no such features are seen
within the inner 200~kpc, in the gas associated with the radio lobes
of 4C\,+24.36. Even if the heating is due to an abundance of buoyant
bubbles, the larger ones should have been evident in the deep X--ray
image. A forthcoming deep X--ray observation (80 ks) at higher
angular resolution, to be obtained by {\it Chandra} during Cycle~9,
will possibly yield a better insight into the presence or absence of
cavities in this system.

\section{Summary and Conclusions}\label{sec:summary}

In this paper we presented a detailed morphological and spectral
analysis, at radio wavelengths, of the WAT radio galaxy 4C\,+24.36,
located at the core of the poor galaxy cluster AWM\,4, based on new
high sensitivity observations at 235~MHz, 327~MHz and 610~MHz, using
the GMRT, and literature and archival data at other frequencies.

The large scale source structure appears bent in a very similar WAT
morphology at all frequencies, and all resolutions afforded by
our GMRT images. We interpret this bending to be the result of the
interplay of ram pressure (driven by the motion of the host galaxy
NGC\,6051) and buoyancy forces. In the framework of this model, we
estimate that the velocity of the galaxy NGC\,6051, with respect to
the cluster, is $\ltsim$ 120 km s$^{-1}$, in the plane of the
sky. Combined with the upper limit on the radial velocity of
NGC\,6051, which is $\ltsim$ 20 km s$^{-1}$ with respect to the mean
radial velocity, out to $0.5\,h^{-1}$~Mpc, of the cluster AWM4
(Koranyi \& Geller 2002), our result indicates that the galaxy is
expected to be relatively at rest in the potential well of the
cluster. 

On the small scale, the radio jets show prominent symmetric
oscillations within the central $\sim$25 kpc from the core. An
appealing possibility to explain these features is the development of
Kelvin--Helmholtz helical instabilities propagating along the
jets. However the resolution of our images is not high enough to
investigate the fine structure of the jets, and to determine the
characteristics of a possible distortion wave. Deep radio observations
at higher resolution, combined with polarisation information, are
necessary to carry out a more detailed analysis of the jet regions.

Combining the new GMRT observations, with data from the
literature at various other frequencies (see Tab.~4), we derived the
broadband radio spectrum of 4C\,+24.36, and found that it is 
well fitted by a power--law with $\alpha=0.82 \pm 0.03$ between 
74 MHz and 10.7~GHz.
From the GMRT images at 235 and 610 MHz, evaluating the spectral 
index distribution at discrete intervals along each jet, we showed 
(Fig.~7) how the spectral index gradually steepens along the jet to 
about $0.9\pm 0.1$ where the jets merge into the lobes, and further 
steadily steepens within the lobes to $\alpha \!>\! 1.5$. 

Assuming equipartition of energy between the relativistic particles and
the ambient magnetic field, and adopting a lower energy cutoff equivalent to
the minimum Lorentz factor of the radiating electrons ($\gamma_{\rm min}$=10,
$E_{\rm min}\sim 5$ MeV), we find that the equipartition value of
the magnetic field varies between 3 to 9$\mu$ G from the lobes to the jets.  
The radiative age for  4C\,+24.36 was found to be around 160~Myr from 
this estimate.

We used the jet to counterjet brightness ratio to constrain the
geometry and velocity of the jet. Assuming a typical value for the
inner jet velocity in WAT radio galaxies, we found that the source is
likely oriented at a large angle ($\sim$ 81$^{\circ}$--88$^{\circ}$)
with respect to the line of sight. This result is consistent with the
high level of symmetry found in the total flux density, extent,
spectral index properties and physical parameters of the lobes and
jets.

Even though the host cluster AWM04 has a hot X--ray emitting ICM
of relaxed morphology and uniform temperature profile out to
at least 160~kpc from the center, its cooling time in the core 
is about 2~Gyr, indicating the presence of significant sources of 
energy injection that prevents the catastrophic cooling expected in 
the core of the cluster. Based on the above values of the jet power 
and the estimated radiative age, the total energy output from the 
radio source falls short by almost two orders of magnitude
of the energy required for the isothermal temperature profile 
obtained from the {\it XMM--Newton} observation of the cluster.
The latter observation enables us to measure the pressure of the hot
gas as a function of radius within the cluster, which is found to be
comparable to the pressure in the jets over most of their lengths, but
higher than the pressure in the lobes by a factor $>$5.  Several
interpretations of this pressure imbalance are discussed. Finally, it
is found that for any reasonable value of gas pressure in the range
between the value in the lobes and that in the jets, if the
central source were to transfer mechanical energy to the ICM, in terms
of the enthalpy of buoyant bubbles of relativistic fluid, it would
require bubbles of tens of kpc in radius, which would have been detected
in the existing XMM-Newton observations of O'Sullivan et al. (2005). We look
forward to a forthcoming deeper {\it Chandra} observation of this system
for further insight into morphological evidence of this energy transfer.

\acknowledgments

We thank the staff of the GMRT for their help during the observations.
GMRT is run by the National Centre for Radio Astrophysics of the Tata 
Institute of Fundamental Research. E.O'S. acknowledges support from NASA
grant NNX06AE90G. J.M. Vrtilek acknowledges support from NASA
grant NNX07AE95G. Basic research in radio astronomy at the NRL is 
supported by 6.1 Base funding. We would like to thank R. Fanti, 
G. Brunetti and M. Markevitch for useful discussion. This research has 
made use of the NASA/IPAC Extragalactic Database (NED) which is operated 
by the Jet Propulsion Laboratory, California Institute of Technology. 
The authors made use of the database CATS (Verkhodanov et al. 1997) of the 
Special Astrophysical Observatory (Russia).




\end{document}